# Affine Geometry, Visual Sensation, and Preference for Symmetry of Things in a Thing

Birgitta Dresp-Langley

ICube Laboratory UMR 7357 CNRS

University of Strasbourg - FRANCE

E-mail: birgitta.dresp@unistra.fr

**Abstract**

Evolution and geometry generate complexity in similar ways. Evolution drives natural selection while geometry may capture the logic of this selection and express it visually, in terms of specific generic properties representing some kind of advantage. Geometry is ideally suited for expressing the logic of evolutionary selection for symmetry, which is found in the shape curves of vein systems and other natural objects such as leaves, cell membranes, or tunnel systems built by ants. The topology and geometry of symmetry is controlled by numerical parameters, which act in analogy with a biological organism's DNA. The introductory part of this paper reviews findings from experiments illustrating the critical role of two-dimensional design parameters and shape symmetry for visual or tactile shape sensation, and for perception-based decision making in populations of experts and non-experts. Thereafter, results from a pilot study on the effects of fractal symmetry, referred to herein as the *symmetry of things in a thing*, on aesthetic judgments and visual preference are presented. In a first experiment (psychophysical scaling procedure), non-expert observers had to rate (scale from 0 to 10) the perceived beauty of a random series of 2D fractal trees with varying degrees of fractal symmetry. In a second experiment (two-alternative forced choice procedure), they had to express their preference for one of two shapes from the series. The shape pairs were presented successively in random order. Results show that the smallest possible fractal deviation from "symmetry of things in a thing" significantly reduces the perceived attractiveness of such shapes. The potential of future studies where different levels of complexity of fractal patterns are weighed against different degrees of symmetry is pointed out in the conclusion.

# Introduction

Brain evolution has produced highly specialized processes which enable us to effectively exploit the geometry of visual perceptual space. Some data suggest that the human brain is equipped with an in-built sense of geometry (e.g. Amir *et al.*, 2012; Amir *et al.*, 2014), which is a key to conscious knowledge about specific object properties and associations between two-dimensional projections and their correlated three-dimensional structures in the real world (e.g. Biederman, 1987; Wilson & Wilkinson, 2002; Pizlo *et al.*, 2010; Li *et al.*, 2013). These associations favour structural regularities and, above all, symmetry (Li *et al.*, 2009; Li *et al.*, 2013). Thus, it is not surprising that symmetry plays an important role in conceptual processes and the design geometry of complex spatial structures, and is abundantly exploited by engineers and architects. The use of the symmetry of curvature, for example, dates back to the dawn of building shelter and vernacular architecture, which relies, by the nature of the materials and construction techniques used, almost entirely on symmetrical curves (Figure 1, left). In the middle ages, descriptive geometry was used for the planning and execution of building projects for which symmetric curves were the reference model, as in the design of arched hallways and corridors (Figure 1, middle). In the last century, the Spanish designer and architect Gaudi exploited the same kind of geometry for the design of the *Sagrada Familia* in Barcelona (Figure 1, right) and many of his other fabulous structures, which can be appreciated by taking a walk through the Güell Park, or by visiting the Güell museum in Barcelona.

FIGURE 1

Gaudi's structures were largely inspired by nature, which abounds with curved shapes and features (see also Ghyka, 1946), and our perception uses these features as cues to shape or object recognition and image interpretation (e.g. Stevens 1981a and b; Foley *et al.*, 2004; Dresp, Silvestri and Motro, 2007; Dresp-Langley 2013, 2015; Mustonen *et al.*, 2015; Strother, Killebrew and Caplovitz, 2015). In biology, curvature guides physical, chemical, and biological processes, such as protein folding, membrane binding, and other biophysical transformations (Grove, 2009). The representation and cognition of curvature ranges from the biochemical level of living organisms capable of sensing this property in their near or distant physical environments (Hatzakis, 2009) to perceptual properties extracted from physical stimuli by the human brain, the ultimate product of evolution. In terms of a mathematical property of the physical world, curve symmetry can be directly linked to affine geometry (see also Gerbino and Zhang, 1991).

*Affine geometry and visual sensation*

In affine geometry, curves derived from circles and ellipses share certain properties, the circle being a particular case of the ellipse. Projective geometry permits generating symmetric curves from ellipses by affinity with concentric circles (Figure 2).

FIGURE 2

Their perception is grounded in biology in the sense that most natural objects can be represented in 2D as symmetrically curved shapes with Euclidean properties of ellipses. Studies comparing between visually perceived curvature by experts in geometry (architects and design engineers) and non-experts (Dresp, Silvestri and Motro, 2006), using symmetric curves derived from concentric circles by affine projection have shown that their perceived magnitude is determined by a single geometric parameter, the curves' *aspect ratio*. The perceptual responses to such curves are independent of both expertise and sensory modality, given that tactile sensing by sighted blindfolded and congenitally blind observers produces the same results (Dresp-Langley, 2013). The symmetry of the curves, however, is a critical factor to these geometry-based perceptual responses (Dresp-Langley, 2015). The *aspect ratio* relates the height (*sagitta*) to the width of a curve, and in symmetric curves of variable size but constant *aspect ratio* directly taken from concentric circles (no projection by affinity), perceived curvature is also constant, in both vision and touch. This observation is directly linked to the phenomenon of scale-invariance in visual curvature discrimination (cf. Whitaker and McGraw, 1998) and in the detection and recognition of shapes in general (cf. Pizlo, 1994).

*Reflection and rotational shape symmetry*

The role of reflection symmetry in visual perception was pointed out by Gestalt psychologists at the beginning of the 20th century (Bahnsen, 1928; Koffka, 1935) as a major factor in shape perception. It refers to specific transformations by transition of points in Euclidean space resulting in mirrored representations. Axial symmetry (e.g.), which results from point-by-point mirroring across an axis ($f(x, y, z) = f(-x, y, z)$), is an important factor in visual recognition (e.g. Braitenberg, 1990; Beck, Pinsk & Kastner, 2005; Tjan & Liu, 2005). Reflection or mirror symmetry is detected fast (Barlow and Reeves, 1979; Wagemans, et al., 1991), in foveal and in peripheral vision (Barrett et al., 1999). Vertical mirror symmetry facilitates face recognition by human (e.g. Thornhill & Gangestad, 1999) and non-human primates (Anderson et al., 2005), and is used by the human visual system as a second-order cue to perceptual grouping (Machilsen et al.,

2009).

Rotational symmetry of shape plays an important role in architecture and design (e;g. Arnheim, 1969). The design of complex modern spatial structures is a domain of contemporary relevance. Visual-spatial experiments on expert architects as well as novices have shown that perceiving the rotational symmetry of partial shapes which constitute the simplest possible *tensegrity* (*tensile integrity*) structure (Figure 4) is an important part of our understanding how they are put together. Only once this symmetry is perceived by the expert or novice, will he/she be able to draw the structure from memory into axonometric or topological reference frames provided to that effect (Silvestri, Motro, Maurin and Dresp-Langley, 2010). Tensegrity structures have inspired current biological models (e.g. Levin, 2002), from the level of single cells to that of the whole human body. They posses what Mandelbrot (1982) called "fractal consistency across spatial scales", or "fractal iterations", like those seen in large trees that appear composed of many smaller trees of the same structure.

*Nature-inspired design and the symmetry of "things in a thing"*
Fractal geometry is also inspired by nature (Mandelbrot, 1982), with its many symmetric visual structures like those found in cells, trees, butterflies and flowers. A fractal may be defined as a complex whole (object or pattern) that has the same structural characteristics as its constituent parts. The structural symmetry that results from fractal iterations may be described as the *symmetry of things in a thing*. The radial symmetry of a sunflower is a choice example of fractal symmetry as it exists in nature. Behavioural studies have shown that various animal species are naturally attracted to two-dimensional representations of objects exhibiting flower-like radial symmetry (Lehrer *et al.*, 1995; Giurfa *et al.*, 1996). In complex 3D fractal trees, single fractals ("things") have a symmetrical counterpart within the whole structure (*the thing*), which may possess radial symmetry, reflection symmetry and manifold rotational symmetries, like many objects in nature (plants, snowflakes, etc.) are bound by both reflection and rotational symmetry, and exhibit multiples of one and the same shape (*things*) repeated in all directions.

Nature-inspired design occupies an important place in contemporary graphic art, and symmetry has been identified as a major defining feature of visual beauty, compositional order, and harmony. Symmetry directly determines aesthetic preferences and the subjectively perceived beauty of two-dimensional visual images and patterns (Eisenman, 1967; Berlyne, 1971; Jacobsen *et al.*, 2002, 2003, 2006; Tinio & Leder, 2009), and symmetrical visual patterns are also more easily remembered and recognized (Deregowski, 1971, 1972; Kayert & Wagemans, 2009) compared with asymmetrical ones. Sabatelli *et al.* (2010) suggested that natural and artistic creative processes rely on common, possibly fractal, transformations. Fractal transformations may describe iterative transitions from simplicity and order (symmetry) to complexity and chaos (asymmetry). Again, fractal trees seem to be a pertinent example here, where simple 2D mirror trees (Figure 3) with reflection and/or radial symmetry open an almost infinite number of possibilities for adding complexity through further transformations leading to complex projections of 3D structures with multiple rotational symmetries (not shown here).

FIGURE 3

Whether nature-inspired fractal design appeals to our senses in the same way as the real objects found in nature remains an open question. However, on the basis of previous findings summarized here above, we may assume that the *symmetry of things in a thing* in fractal design plays a decisive role in our perception of their aesthetic content and thereby influence certain preference judgments. Given the multiple levels of complexity of fractal objects, trying to address this question requires starting with simple examples. For this pilot study here, we created a series of fractal mirror trees based on geometric transformations as shown in Figures 2 and 3. In two psychophysical experiments, one using a subjective aesthetic rating procedure, the other a preference judgment design, we tested whether the subjective attractiveness of such trees is affected by different degrees of violation of symmetry, from an almost imperceptible lack of mirror detail to massive asymmetry.

**Materials and methods**

The experiments were conducted in accordance with the Declaration of Helsinki (1964) and with the full approval of the corresponding author's institutional (CNRS) ethics committee. Informed written consent was obtained from each of the participants. Experimental sessions were organized following conditions of randomized, trial-by-trial free image viewing using a computer with a keyboard and a high resolution monitor. 15 mirror tree images were generated using a comprehensive vector graphics environment (Adobe Illustrator CC) and computer shape library.

*Subjects*

30 observers, ranging in age between 25 and 70 and unaware of the hypotheses of the study, participated in the experiments. All subjects had normal or corrected-to-normal visual acuity.

*Stimuli*

The stimuli for the two experiments were generated on the basis of 15 images of fractal trees (Figure 5) drawn in a vector graphics environment (Adobe illustrator CC) using simple principles of 2D geometry, as shown here above in Figure 3. Five of these images (Figure 4, top row) were mirror trees with vertical reflection symmetry and perfect *symmetry of things in a thing*. Five of them (Figure 4, middle row) were imperfect mirror trees in the sense that their vertical reflection symmetry excluded one of the elementary parts, which was not mirrored on the right side of the tree. In the remaining five, asymmetrical images Figure 4, bottom row), elementary shapes "growing" on the branches of the left side of the trees were not mirrored on the right side. The luminance contrast between figures and backgrounds was constant in the 15 images (same RGB (200, 200, 200) for all figures, same RGB (20, 20, 20) for all backgrounds). The height of a fractal tree on the screen was 10 cm, the widest lateral expansion in the vertical direction of any given tree was 4 cm.

FIGURE 4

*Task instructions*

In the aesthetic rating experiment, subjects were instructed to rate the beauty of each of the fifteen individual images on a subjective psychophysical scale from 0 (zero) for "very ugly" to ten (10) for "very beautiful". In the preference judgment experiment, subjects were instructed to indicate whether they spontaneously preferred the left or the right of an image pair. Hitting the response key initiated the next image pair. Half of the subjects started with the rating experiment, the other half with the preference judgment experiment.

*Procedure*

Subjects were seated at a distance of 1 meter from the screen and looked at the center of the screen. The images were displayed centrally and presented in random order. In the aesthetic rating experiment, each of the 15 images was presented once to each of the 30 subjects. In the preference judgment experiment, each image from a group of five was paired with its counterpart from the two other groups of five. Their spatial position in a pair (left/right) was counterbalanced. This yielded 20 image pairs that were presented twice each to each of the 30 subjects in an individual session. Individual responses were coded and written into text files for further processing. Inter-stimulation intervals were observer controlled. They typically varied from one to three seconds, depending on the observer, who initiated the next image presentation by striking a given response key on the computer keyboard.

**Results**

The raw data from the two experiments were analyzed using *Systat 11*. Data plots showing medians and variances of the rating distributions were generated. Means and their standard errors of the subjective aesthetic ratings and the total number of "preferred" responses from the preference judgment task were plotted for comparison between figure types. One-way analyses of variance testing for statistical significance of differences in means observed for the three figure types: 'symmetrical', 'single detail missing on right' and 'asymmetrical' were performed.

*Subjective aesthetic ratings*

The medians and variance of the subjective aesthetic ratings between zero and ten produced by the 30 subjects in response to the 15 images were plotted as a function of the three-level figure type factor (Figure 5). With five figures per factor level and 30 individual ratings per figure, we have a total of 150 observations for each level of this factor, and a total of 450 observations. The distribution of observations satisfies criteria of normality and equality of variance for further parametric testing, outliers were not removed from the dataset. One-way ANOVA signaled a significant effect of figure type on raw data for subjective beauty ratings ($F(2, 449)=79.47$; $p<.001$). The differences between the means, plotted here in terms of the average subjective rating and its standard error for each figure type (Figure 6), reveal that perfectly symmetrical figures score higher for subjective beauty than figures with a detail missing ($t(1, 149)=7.15$; $p<.001$), *post hoc* Holm-Sidak comparison), and that figures with a detail missing score higher than asymmetrical figures ($t(1, 149)=5.42$; $p<.001$), *post hoc* Holm-Sidak comparison). The largest difference in average aesthetic ratings is observed between symmetrical and asymmetrical figures ($t(1, 149)=12.57$; $p<.001$, *post hoc* Holm-Sidak comparison). When average beauty ratings are plotted as a function of the individual figures (Figure 7), we see that none of the three figure types produced an average score in the extremes ("very beautiful" or "very ugly"). The five symmetrical ones (1 to 5 on the x-axis) produced average ratings between '5' and '8', the five with a small detail missing on the right (6 to 10 on the x-axis) produced average scores between '4' and '6', and the five asymmetrical figures (11 to 15 on the x-axis) scored between '3' and '4' on average.

FIGURE 5

FIGURE 6

FIGURE 7

*Preference judgments*

The total number of times each figure of the 15 was chosen as "preferred" in a pair of images in the preference judgment task was counted. One-way ANOVA on the total number of preferences for a figure of each type (N=5 per factor level) signaled a significant effect of figure type on preference ($F(2, 14)=368.12$; $p<.001$). The differences between means, plotted here in terms of the average number of "preferred" and its standard error for each figure type (Figure 8), reveal that perfectly symmetrical figures yield larger preferences than figures with a detail missing ($t(1, 4)=19.00$; $p<.001$), *post hoc* Holm-Sidak comparison), and that figures with a detail missing yield larger preferences than asymmetrical figures ($t(1, 4)=7.28$; $p<.001$), *post hoc* Holm-Sidak comparison). The largest difference in number of "preferred" is observed between symmetrical and asymmetrical figures ($t(1, 4)=26.28$; $p<.001$, *post hoc* Holm-Sidak comparison). When the total number of "preferred" responses is plotted as a function of the 15 individual figures (Figure 9), we see that the five symmetrical figures (1 to 5 on the x-axis) produced almost identical high-preference totals, while the other figures produced more variable ones in the lower preference range.

FIGURE 8

FIGURE 9

**Discussion**

As illustrated by examples from the introduction here above, shape sensation and perception can be related to affine design geometry (e.g. Bahnsen, 1928; Koffka, 1935; Braitenberg, 1990; Gerbino and Zhang, 1991; Dresp-Langley, 2015). Similarly, the topology and geometry of fractal objects may be controlled by a few simple geometric parameters, as in the fractal mirror trees that were used as stimuli here. The term "fractal" was first introduced by Mandelbrot (1982) based on the meaning "broken" or "fractured" (*fractus*), with reference to geometric patterns existing in nature. The findings from this study here show that the smallest "fractal" deviation from perfect *symmetry of things in a thing* in basic mirror trees (any computer shape library can generate them) with vertical reflection symmetry when no fractals are removed, significantly diminishes subjectively perceived beauty and visual preference. These results confirm previous observations from aesthetic perception studies using different two-dimensional configurations (Eisenman, 1967; Berlyne, 1971; Jacobsen *et al.*, 2002, 2003, 2006; Tinio & Leder, 2009). Perfectly symmetrical trees also produced the strongest consensual results, for both subjective aesthetic ratings and visual preferences, while the ones with a small detail missing and the asymmetrical trees produced more disparate data, indicating higher uncertainty (i.e. less confidence) in the subjects' perceptual responses.

In nature, it is indeed difficult to find things which do not have at least one axis of mirror or reflection symmetry, such as palm trees and sunflowers or broccoli and snowflakes (cf. Mandelbrot, 1975), for example. Also, most human beings are basically symmetric around the vertical axis when standing up, and it is therefore almost unsurprising that our aesthetic preferences would mostly go for symmetrical objects (see also Tinio & Leder, 2009, on massive familiarization). However, results from earlier studies (Eisenman & Gellens, 1968) lead to suggest that things may not be that simple when complexity and symmetry are weighed against each other, and when socio-cultural factors are brought into the equation. Personality and creativity (Eisenman and Rappaport, 1967; Arnheim, 1969; Cook and Furnham, 2012) have been identified as two such variables, and highly creative individuals may have a stronger tendency to prefer asymmetrical objects, especially when these exhibit high levels of complexity, as in the case of fractal objects with multiple rotational symmetries, for example. As pointed out previously (Sabatelli *et al.*, 2010), symmetry and asymmetry coexist in many natural and human processes, and the critical role of symmetry in art has been well demonstrated; the complementary role of asymmetry maybe less . Fractal objects offer new perspectives for research on complementary aspects of symmetry and asymmetry in processes of increasing complexity, including processes of visual perception.

Fractals are different from other geometric figures because of the way in which they scale across multiple iterations, yielding increasingly complex repetitive structures which are symmetrical by nature. Fractal symmetry is also referred to as *expanding symmetry* or *evolving symmetry*, especially if replication is exactly the same at every scale, as in a detailed pattern that repeats itself across multiple fractal iterations. For the visual scientist, this opens many perspectives as it permits the finely controlled manipulation of each and every shape detail in a given configuration and thereby allows to create visual stimuli where variations in complexity and symmetry can be effectively weighed against each other in further studies.

**Conclusion**

The visual attractiveness of 2D fractal design shapes closely depends on the *symmetry of things in a thing* in configurations with simple geometry, as shown in this pilot study here on the example of a few very basic fractal mirror-trees. In these simple displays, the smallest "fractal" deviation from a perfect *symmetry of things in a thing* is shown to have significantly negative effects on subjectively perceived beauty and preference judgments. These findings are to encourage further studies, using more sophisticated fractal design objects with increasingly large number of fractal iterations, producing more and more complex 2D mirror designs and shapes with increasingly multiple rotational symmetry in 3D. Such design objects are ideally suited for a numerically controlled manipulation of the smallest of details in the *symmetry of things in a thing*, perfectly tailored for investigating complex interactions between symmetry and complexity in their effects on visual sensation and aesthetic perception.

**Figure captions**

FIGURE 1

The importance of the symmetry of curves for human endeavour dates back to the dawn of building shelter and to vernacular architecture (left). Symmetric curve geometry is currently used in contemporary free-form architecture (middle), which has been much inspired by the Spanish architect Gaudi, who largely exploited symmetry of curvature for the design of the hall and archways of the *Sagrada Familia* in Barcelona (right).

FIGURE 2

Projective geometry permits generating symmetric curves from ellipses by affinity with concentric circles .Their perception is grounded in biology in the sense that most natural objects can be represented in images as symmetrically curved shapes with the Euclidean properties of ellipses. Symmetric curves yield visual and tactile sensations of curvature which increase exponentially with the *aspect ratio* of the curves (e.g. Dresp-Langley, 2013; 2015)

FIGURE 3

Fractal geometry and affine geometry share principles of projection in Euclidean space. Fractal trees, inspired by nature (Mandelbrot, 1982), may be defined as complex wholes where every part repeats itself across multiple fractal iterations, producing *symmetry of things in a thing*. In the simple fractal mirror-tree shown here, concentric circles are the mathematical basis for describing structural regularities with vertical reflection (mirror) symmetry, which has been identified as a major determinant of the visual attractiveness of image configurations (e.g. Eisenman, 1967).

FIGURE 4

Stimuli from the aesthetic rating and visual preference experiments described herein. Fifteen images of fractal mirror trees were designed using some of the principles of transformation shown in Figures 2 and 3. The first five trees (top) possess perfect *symmetry of things in a thing* across the vertical axis. In the next set of five (middle), the smallest of fractal details is missing on the right. The remaining five trees (bottom) are asymmetrical.

FIGURE 5

Box-plots showing medians and variance of aesthetic ratings (on a scale between zero and ten) produced by 30 naive subjects in response to the 15 images from Figure 4, plotted as a function of figure type.

FIGURE 6

Average aesthetic ratings and their standard errors, plotted for each figure type.

FIGURE 7

Average aesthetic ratings plotted as a function of the 15 individual images.

FIGURE 8

Average number of "preferred" responses and their standard errors, plotted for each figure type.

FIGURE 9

Total number of "preferred" responses plotted as a function of the 15 individual images.

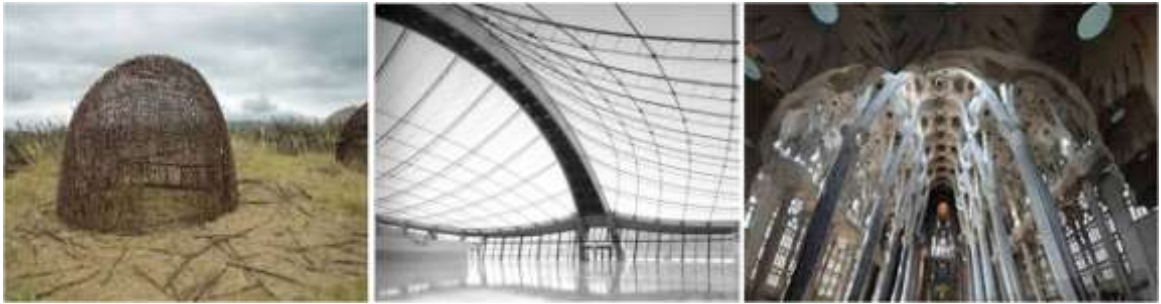

FIGURE 1

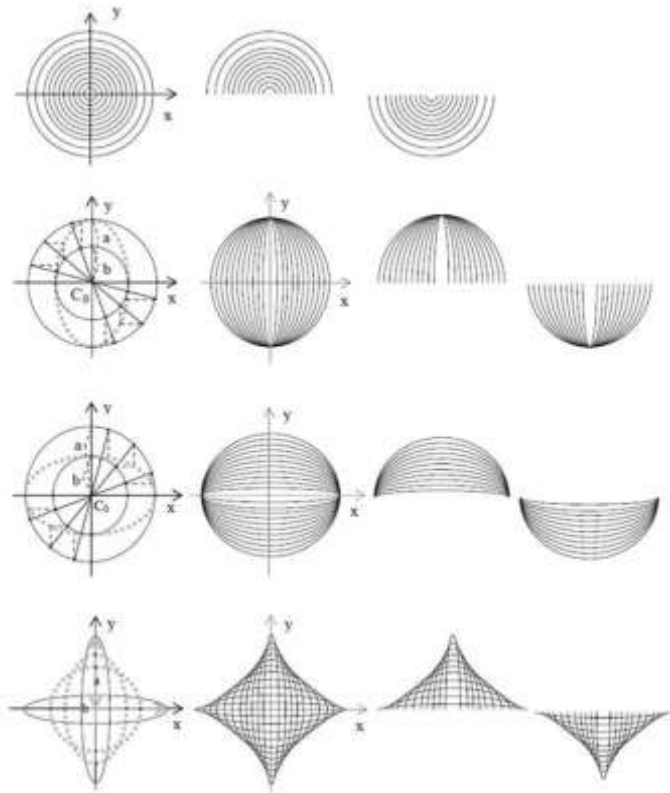

FIGURE 2

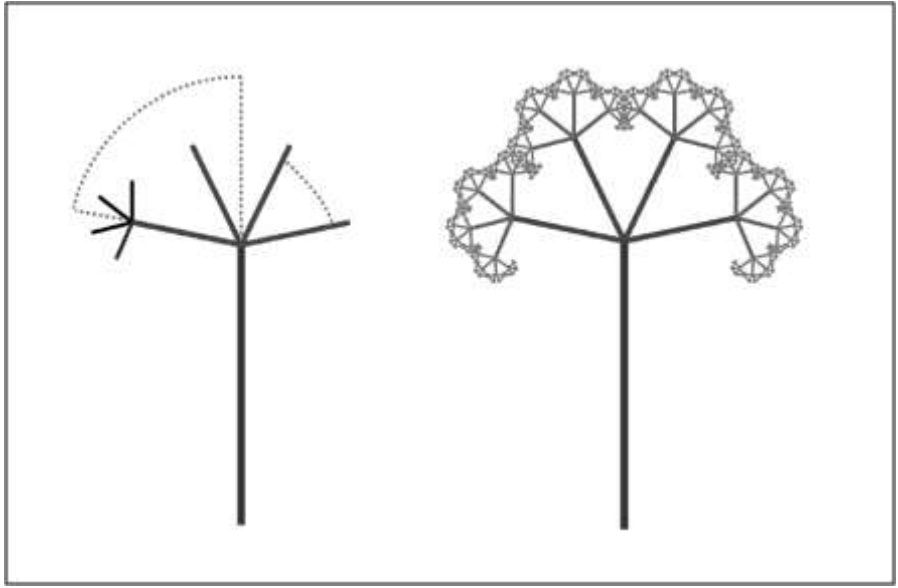

FIGURE 3

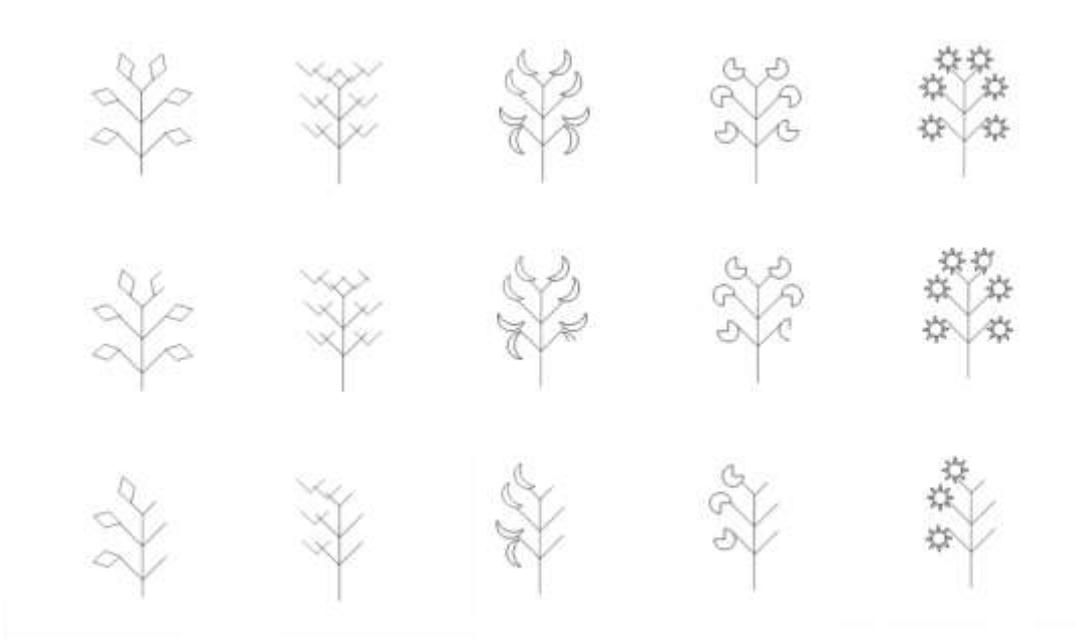

FIGURE 4

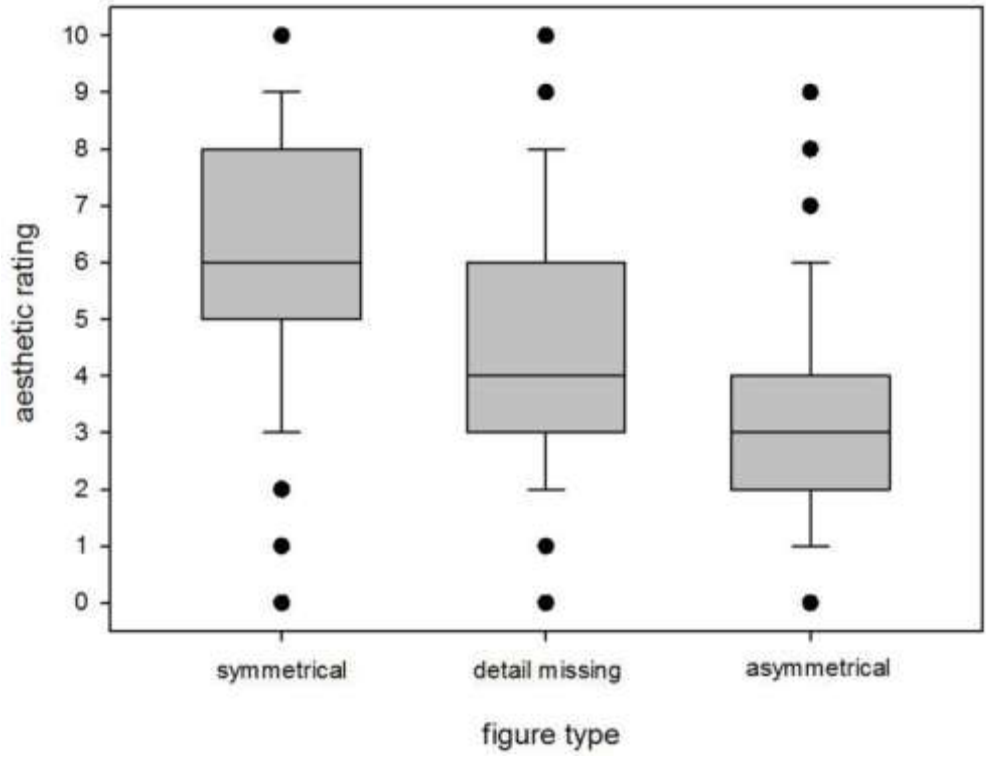

FIGURE 5

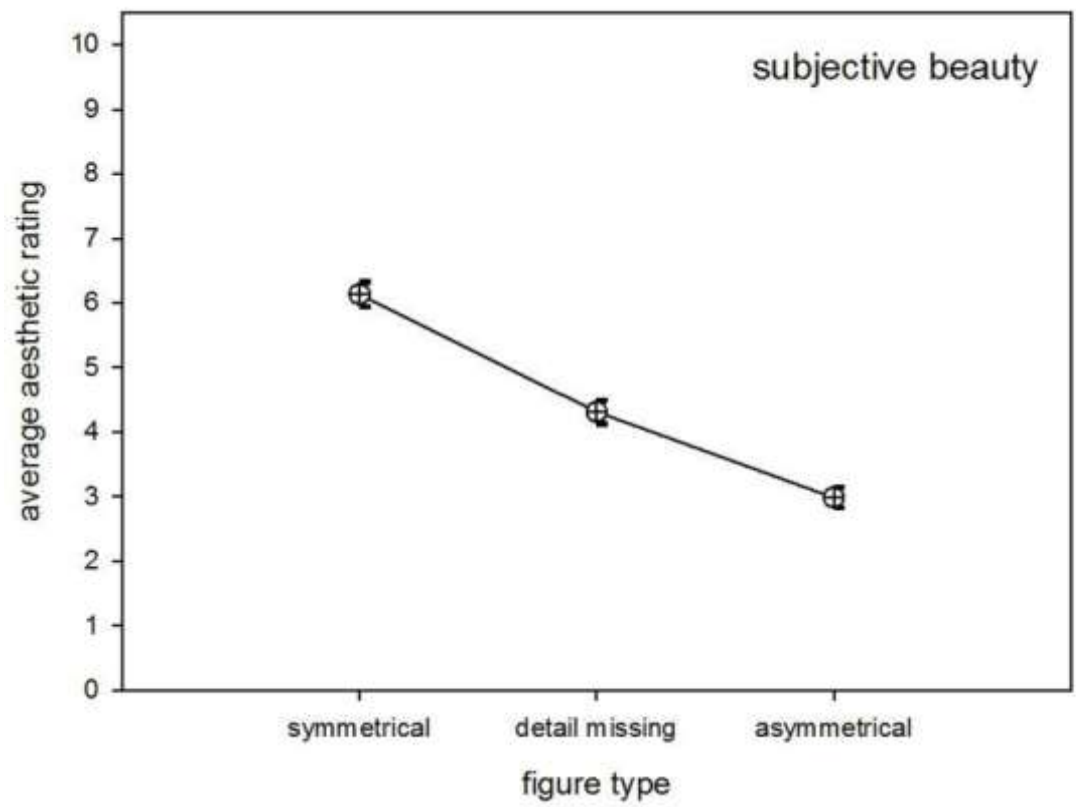

FIGURE 6

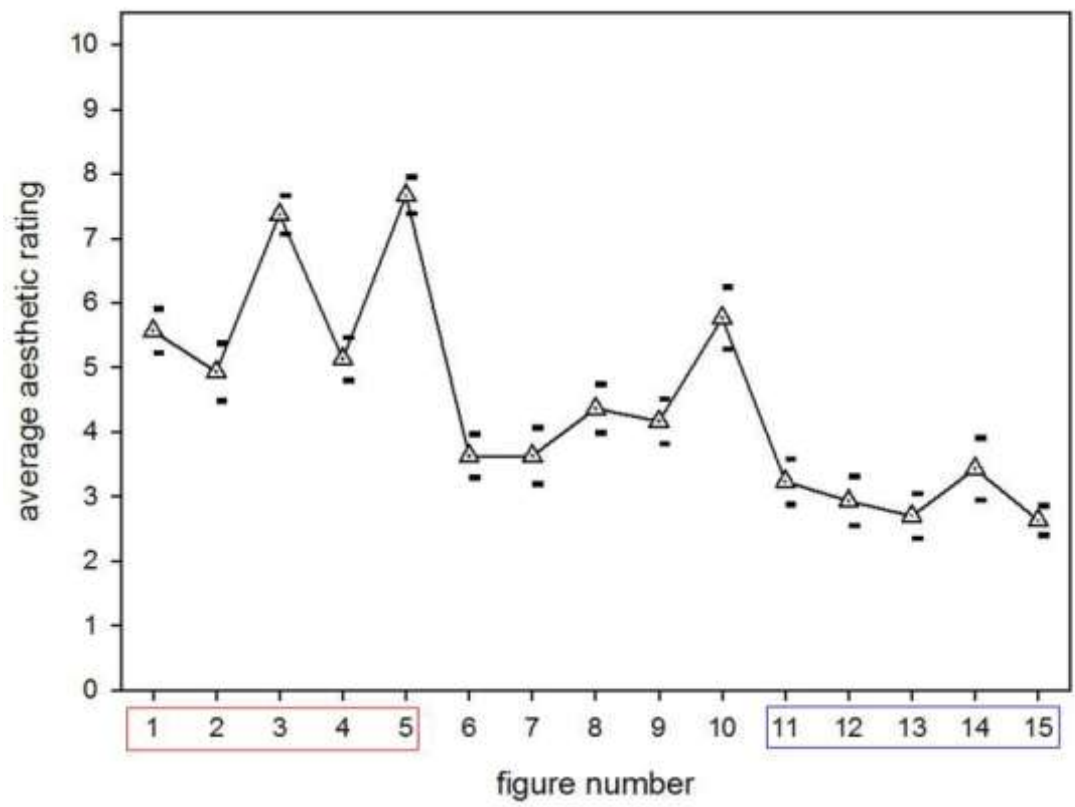

FIGURE 7

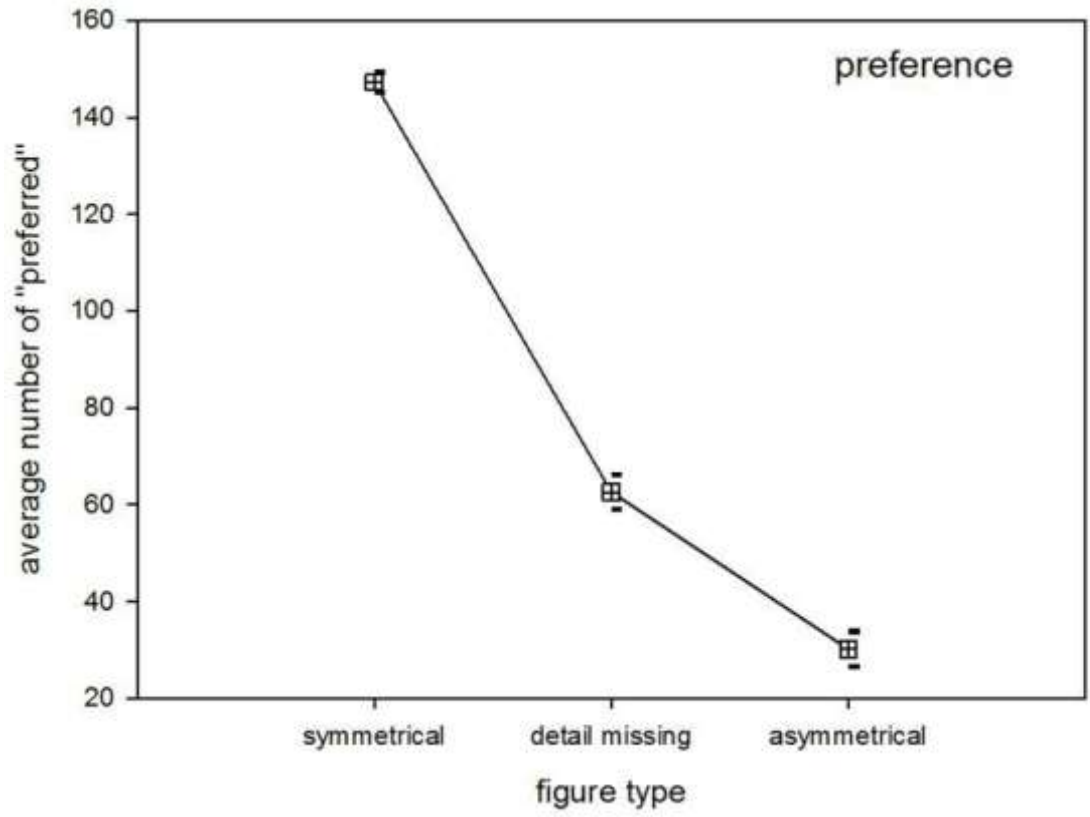

FIGURE 8

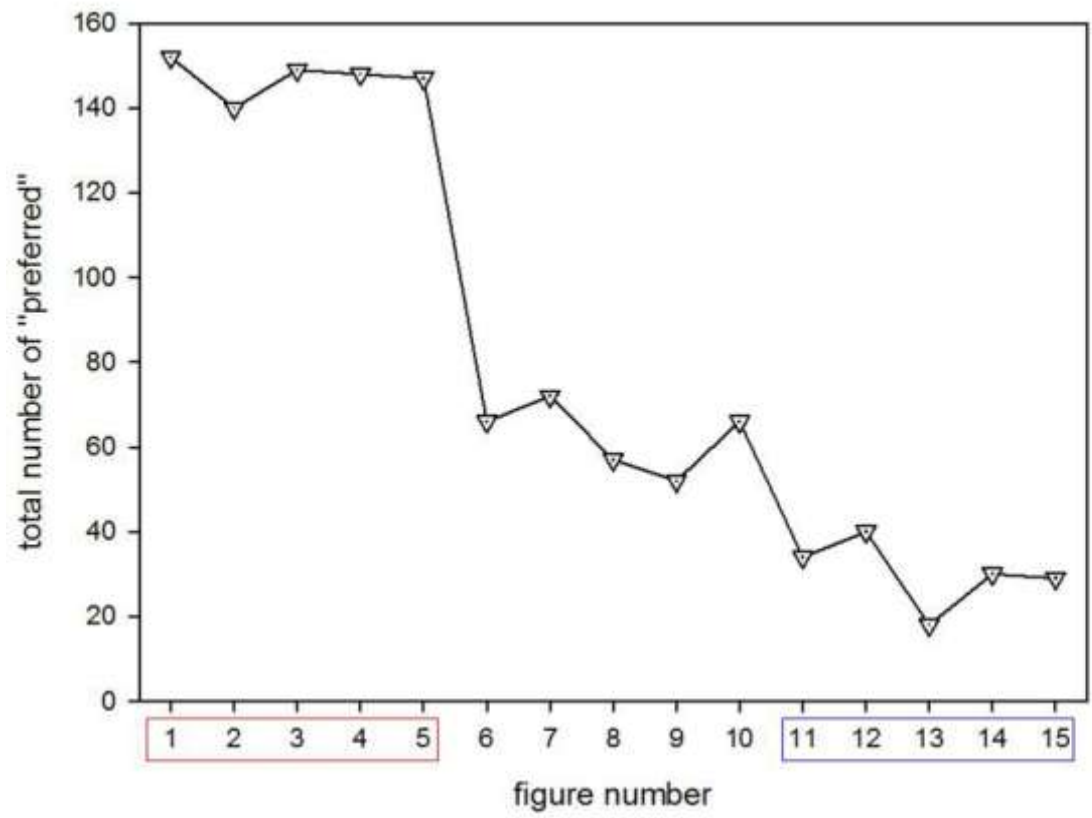

FIGURE 9